\documentclass[twocolumn,prl]{revtex4}
\usepackage{amssymb}
\usepackage{graphicx}

\begin{document}

%\begin{frontmatter}

\title{Encapsulating C$_{59}$N azafullerene derivatives inside single-wall
carbon nanotubes}
\author{$^{1}$F. Simon$^{\ast }$}
\author{$^{1}$H. Kuzmany}
\author{$^{2}$J. Bernardi}
\author{$^{3}$F. Hauke}
\author{$^{3}$A. Hirsch}
\affiliation{$^{1}$ Institut f\"{u}r Materialphysik, Universit\"{a}t
Wien, Strudlhofgasse 4, A-1090 Wien, Austria} \affiliation{$^{2}$
University Service Centre for Transmission Electron Microscopy
(USTEM), Technische Universit\"{a}t Wien, Wiedner Hauptstrasse 8 -
10 / 052, A-1040 Wien, Austria}
\affiliation{$^{3}$ Institut f\"{u}r Organische Chemie der Friedrich Alexander Universit%
\"{a}t Erlangen-N\"{u}rnberg. Henkestrasse 42, D - 91054 Erlangen}

\begin{abstract}
Filling of single-wall carbon nanotubes with C$_{59}$N azafullerene
derivatives is reported from toluene solvent at ambient temperature. The
filling is characterized by high resolution transmission electron microscopy
and Raman spectroscopy. The filling efficiency is the same as for C$_{60}$
fullerenes and the tube-azafullerene interaction is similar to the tube-C$%
_{60}$ interaction. Vacuum annealing of the encapsulated
azafullerene results in the growth of inner tubes, however no
spectroscopic signature of nitrogen built in the inner walls is
detected.
\end{abstract}

%\begin{keyword}
%Carbon nanotubes \sep Fullerene \sep Electron microscopy \sep Raman spectroscopy
%\end{keyword}

%\end{frontmatter}

\maketitle

%\today

% main text

\section{Introduction}

The hollow space inside single-wall carbon nanotubes (SWCNTs) have attracted
considerable attention recently. The field was opened by the discovery of C$%
_{60}$ fullerenes encapsulated inside SWCNTs, the peapods
\cite{SmithNAT}. Later it was found that chemical reactions can take
place inside the tube such as charging induced polymerization
\cite{PichlerPRL2001} or fusion of the C$_{60}$ spheres to inner
tubes \cite{LuzziCPL2000,BandowCPL2001}. The resulting inner tubes
in the double-wall carbon nanotubes (DWCNTs) were shown to be
exceptionally defect free \cite{PfeifferPRL2003} which apostrophizes
the inside the of the tube as "nano clean-room chemical reactor". We
recently found that chemical reactions are not restricted to
fullerenes but organic solvents such as benzene or toluene can also
participate \cite{Simonunpublished}. The breakthrough to further
explore the in-the-tube chemistry was the discovery of encapsulating
fullerenes or fullerene derivatives at ambient temperatures \cite%
{Monthioux2003,YudasakaCPL,SimonCPL2004,Monthioux2004,BriggsJMC}.
Conventional peapod synthesis involves
heating the sample above 400-500 $^{\circ }$C \cite%
{LuzziCPL1999,KatauraSM2001}, which most fullerene derivatives do not
tolerate. C$_{59} $N, the on-ball nitrogen doped modification of fullerenes,
has a rich chemistry due to its enhanced reactivity as compared to pristine
fullerenes and can be synthesized in macroscopic amounts chemically \cite%
{WudlReview,HirschC59NReview}. The electronic state of C$_{59}$N and its
derivatives is strongly modified compared to C$_{60}$ \cite{PichlerPRL1997}.
Encapsulating azafullerene peapods would be advantageous as they are
expected to go preferably inside the SWCNTs similarly to all-carbon
fullerenes, however their sizeable dipole moment adds a further degree of
freedom for their applications such as e.g. ambipolar transistor \cite%
{ShinoharaAPL2002}. In addition, the presence of the nitrogen enables to
explore the in-the-tube chemistry with heteroatoms.

Here, we present the encapsulation of azafullerene derivatives inside
SWCNTs. We use a low temperature synthesis method at ambient conditions. The
encapsulation is proven by high-resolution transmission microscopy and Raman
spectroscopy. The latter method shows that azafullerenes enter the tube with
the same efficiency as C$_{60}$. Inner tubes grown from the azafullerene
adduct are spectroscopically identical to all-carbon inner tubes.

\section{Experimental}

\begin{figure}[tbp]
\includegraphics[width=0.8\hsize]{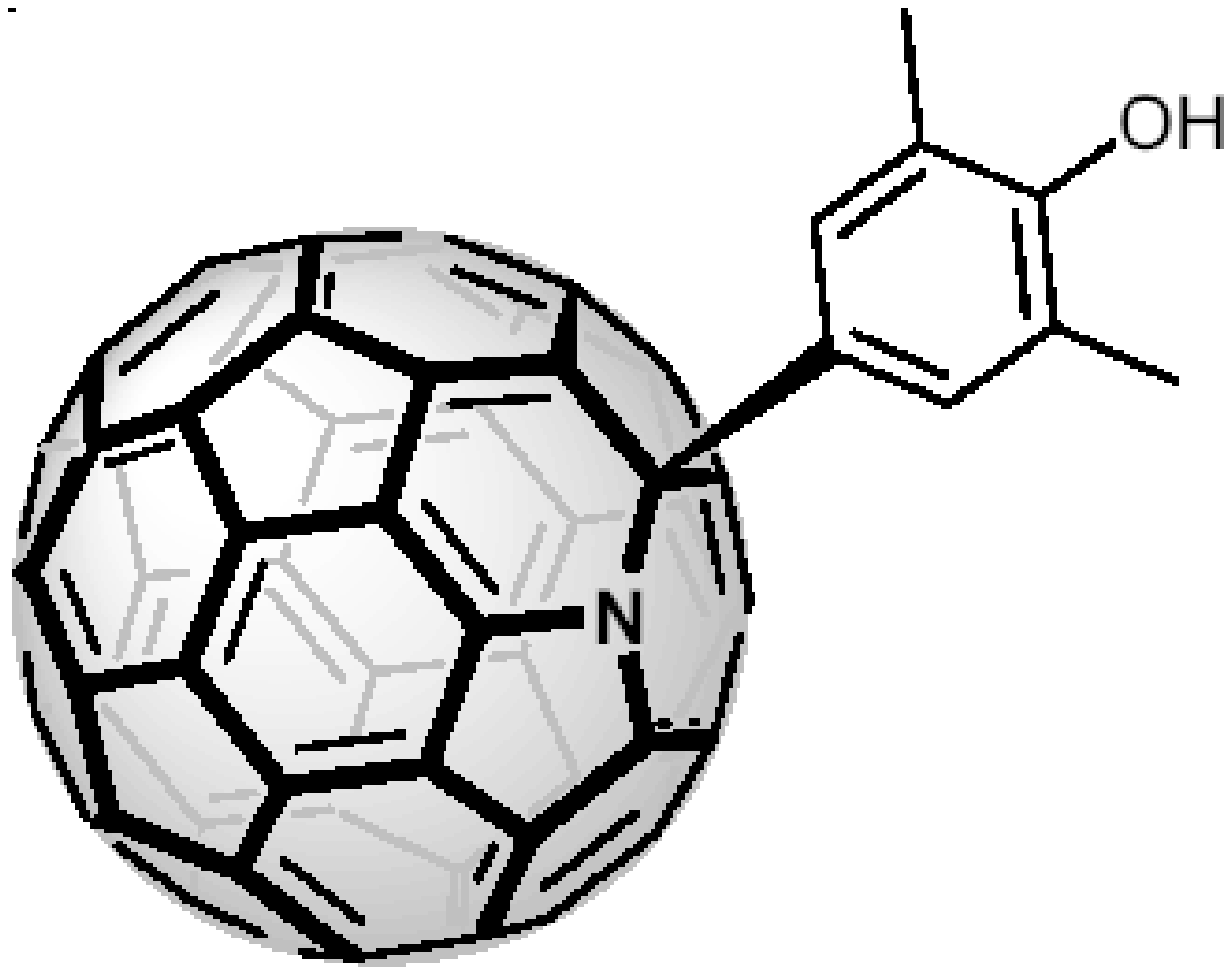}
\caption{Schematic structure of the
4-Hydroxy-3,5-dimethyl-phenyl-hydroazafullerene.} \label{molecule}
\end{figure}

Commercial SWCNT (Nanocarblab, Moscow, Russia) was used in the current
study. The SWCNT material is prepared by the arc-discharge method and is
purified to 50 wt\% by the manufacturer. The mean value, $d$ = 1.40 nm, and
the variance, $\sigma $ = 0.1 nm, of the tube diameters were determined from
multi-laser Raman measurements \cite{KuzmanyEPJB}. The starting
azafullerene, (C$_{59}$N)$_{2}$, was prepared according to literature
procedures \cite{WudlReview}. The
4-Hydroxy-3,5-dimethyl-phenyl-hydroazafullerene (C$_{59}$N-der in the
following) was prepared from 60 mg (41.66
%TCIMACRO{\U{b5}}%
%BeginExpansion
$\mu$%
%EndExpansion
mol) (C$_{59}$N)$_{2}$ and 135 mg (0.7 mmol, 10 eq.) p-Toluenesulfonic acid
dissolved in 100 ml 1,2-Dichlorobenzene. 43 mg (351 $\mu $mol, 5 eq.)
2,6-dimethylpenol was added to this solution. The reaction mixture was
heated to 150 $^{\circ }$C for 15 min while passing a constant stream of air
through the solution. The formed product was isolated by flash
chromatography using toluene as eluent. The product was precipitated from CS$%
_{2}$/pentane, washed three times with pentane and dried in high
vacuum and its molecular structure is shown in Fig. \ref{molecule}.
The material was characterized by $^{1}$H and $^{13}$C NMR and mass
spectroscopy. Fullerene encapsulation was performed with the
modification of the low temperature solvent method
\cite{SimonCPL2004}: open SWCNTs were added to 1 mg/1 ml
fullerene-toluene solutions and sonicated for 1 hour in an
ultrasonic bath (ELMA T460H, 35 kHz, 600 W power). The resulting
material was filtered from the solvent, re-suspended in excess
toluene to remove non-encapsulated fullerenes and re-filtered. Raman
spectroscopy was performed on the bucky-papers. The material was
then vacuum annealed at 1250 $^{\circ }$C for 2 hours for the growth
of inner tubes from the encapsulated material following Refs.
\cite{LuzziCPL2000,BandowCPL2001}.

High resolution transmission electron microscopic (HR-TEM) studies were
performed on a TECNAI F20 field emission microscope at 120 kV. The use of
this voltage combined with low exposition times of 1 s allows to take good
quality pictures without observable irradiation damage to the sample. HR-TEM
samples were prepared from a suspension of the peapod material in
N,N-Dimethylformamide.

Raman spectroscopy was studied on a Dilor xy triple spectrometer in the
488-676 nm range with an Ar-Kr laser at room temperature. We used Raman
spectroscopy to characterize the diameter distribution of the SWCNTs, to
determine the concentration of encapsulated fullerenes, and to study the
resulting DWCNT samples.\smallskip\

%\begin{figure}[tbp]
%\includegraphics[width=0.8\hsize]{molecule.eps}
%\caption{Schematic structure of the
%4-Hydroxy-3,5-dimethyl-phenyl-hydroazafullerene.} \label{molecule}
%\end{figure}

\section{Results and discussion}

\begin{figure}[tbp]
\includegraphics[width=1\hsize]{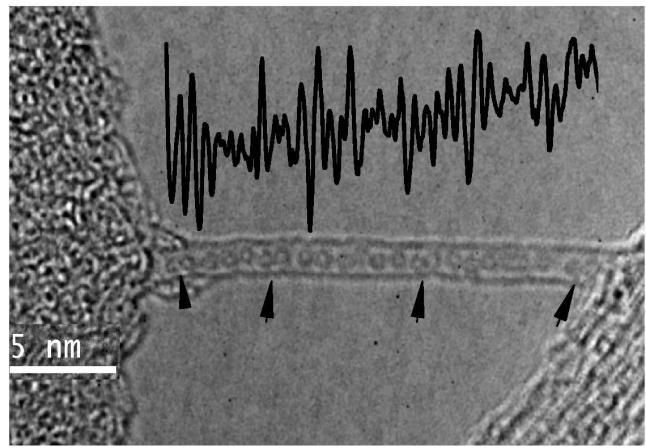}
\caption{HR-TEM micrograph of C$_{59}$N-der encapsulated inside
SWCNTs. The
solid line shows a cross-section profile of the micrograph. Arrows indicate C%
$_{59}$N-der pairs whose centers are only 0.7-0.8 nm apart.}
\label{TEM}
\end{figure}

In Figure \ref{TEM}., we show a HR-TEM\ micrograph of the
C$_{59}$N-der encapsulated inside SWCNTs. HR-TEM shows an abundant
filling of the tubes with the azafullerene, however it does not
provide a quantitative measurement on the filling efficiency that is
determined from Raman spectroscopy. A cross section profile through
the center of the encapsulated azafullerenes enables to determine
their separation as the low and high values of the profile indicate
bright and dark parts, respectively. Interestingly, we found 0.7-0.8
nm separation for some C$_{59}$N-der pairs (indicated by arrows in
Fig. \ref{TEM}) in contrast to the $\sim 1$ nm
separation that is observed for encapsulated C$_{60}$ peapods \cite%
{SmithNAT,HiraharaPRB}. Although the data does not allow to
determine the precise configuration of the C$_{59}$N-der pairs,
neither gives an accurate measure of their separation, it might be
attributed to the presence of the strongly polar side-group of
C$_{59}$N-der.

\begin{figure}[tbp]
\includegraphics[width=1\hsize]{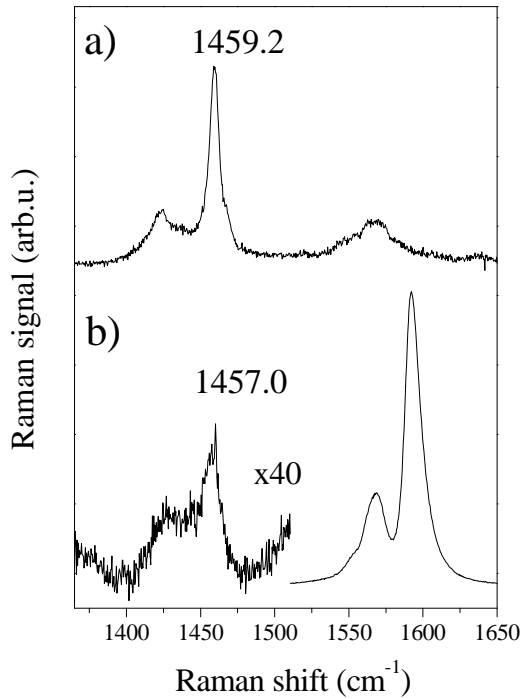}
\caption{Tangential mode of the Raman spectra of the C$_{59}$N-der
before (a) and after encapsulation (b) excited with a 488 nm laser.
Labels mark the position of the strongest C$_{59}$N-der mode in the
two samples. The Raman G-mode of SWCNT dominates the peapod spectrum
in the 1550-1650 cm$^{-1}$ range. } \label{peapodspectra}
\end{figure}

In Figure \ref{peapodspectra}., we show the Raman spectra of the pristine
and encapsulated C$_{59}$N-der. The Raman spectra of the peapod sample
(lower curve in Fig. \ref{peapodspectra}) in the plotted frequency range
consist of the SWCNT G modes around 1550 cm$^{-1}$ and additional lines
related to the Raman active modes of the encapsulated azafullerene
derivative \cite{PichlerPRL2001}. The major Raman modes of the pristine C$%
_{59}$N-der are similar to those of the (C$_{59}$N)$_{2}$ dimer \cite%
{KuzmanyPRB1999}. Here, we focus on the strongest mode that is observed at
1459.2 cm$^{-1}$. This mode is derived from the C$_{60}$ A$_{g}$(2) mode and
is downshifted to 1457 cm$^{-1}$ after the encapsulation procedure. The 2.2
cm$^{-1}$ downshift proves the encapsulation of the molecule inside the
SWCNT. When encapsulated inside SWCNTs, the corresponding A$_{g}$(2) mode of
C$_{60}$ downshifts with 3 cm$^{-1}$ , which is assigned to the softening of
the C$_{60}$ A$_{g}$(2) vibrational mode due to the interaction between the
ball and the SWCNT wall \cite{PichlerPRL2001}. The slight difference between
the downshift for the azafullerene and for the C$_{60}$ peapods might be
attributed to the different structure of the two molecules. The
encapsulation also manifests in a line broadening: the main component of the
1457 cm$^{-1}$ mode is broadened from 4.5 cm$^{-1}$ FWHM in the pristine
material to 10 cm$^{-1}$ FWHM in the encapsulated one. This is similar to
the values found for encapsulated C$_{60}$ \cite{PichlerPRL2001}.

The integrated intensity of the observed A$_{g}$(2) derived mode of the C$%
_{59}$N is approximately 5 times larger than that of a C$_{60}$ peapod
prepared identically when normalized by the SWCNT G mode intensity. This,
however, can not be used to measure the encapsulation efficiency as Raman
intensities depend on the strength of the Raman resonance enhancement and
the Raman scattering matrix elements \cite{KuzmanyBook}. For C$_{60}$
peapods the Raman signal was calibrated with independent and carbon number
sensitive measurements: EELS studies gave the total number of C$_{60}$
related and non-C$_{60}$ related carbons \cite{LiuPRB2002} and the mass of
encapsulated C$_{60}$s was determined from NMR studies using $^{13}$C
enriched fullerenes \cite{SimonPRL2005}. In the current case, neither
methods can be employed and we determined the filling efficiency for the
azafullerene by encapsulating a mixture of the azafullerene and C$_{60}$.

\begin{figure}[tbp]
\includegraphics[width=0.9\hsize]{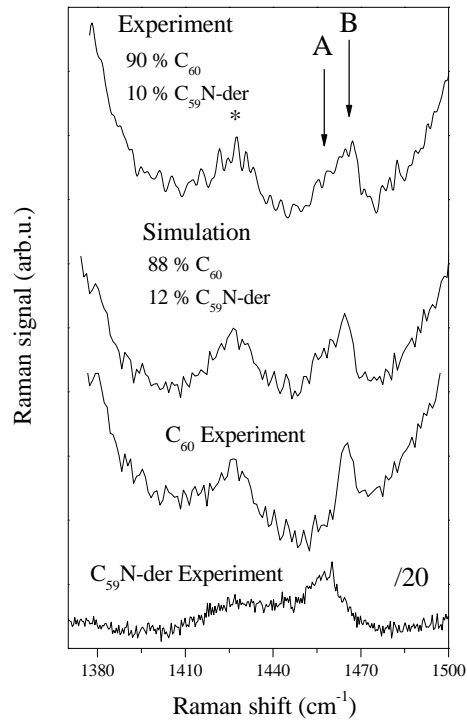}
\caption{Raman spectra of the encapsulated C$_{59}$N-der:C$_{60}$
mixture at the 488 nm laser excitation. The spectra for the
C$_{59}$N-der and C$_{60}$ peapods is shown together with their
weighted sum as explained in the text. A and B mark the componenents
coming nominally from the superposing two phases. The asterisk marks
a mode that is present in the pristine SWCNT material. Note the
different scale for the C$_{59}$N-der peapod material.}
\label{mixedpeapodspectra}
\end{figure}

In Fig. \ref{mixedpeapodspectra}., we show the Raman spectra of the
encapsulated C$_{59}$N-der:C$_{60}$ mixture with weight ratios of 1:9 in the
starting solvent. The Raman spectrum of the encapsulated mixture was
simulated with a weighted sum of the separately recorded spectra for
encapsulated C$_{59}$N-der and C$_{60}$. The best agreement between the
simulated and the experimental spectra is for a C$_{59}$N-der content of
0.12(2). This value is close to the expected value of 0.1 and it proves that
the azafullerene enters the tubes with the same efficiency as C$_{60}$. The
filling efficiency for C$_{60}$ using the toluene stirring method was found
to yield 48(1) \% filling of the total available volume by comparing the
Raman signal with the previous calibration \cite{KuzmanyAPA}. This
corresponds to $\sim $6 \% fraction of the total sample mass as the 100 \%
encapsulation corresponds to 13 \% mass fraction of the encapsulated
fullerenes \cite{SimonPRL2005}.

\begin{figure}[tbp]
\includegraphics[width=0.9\hsize]{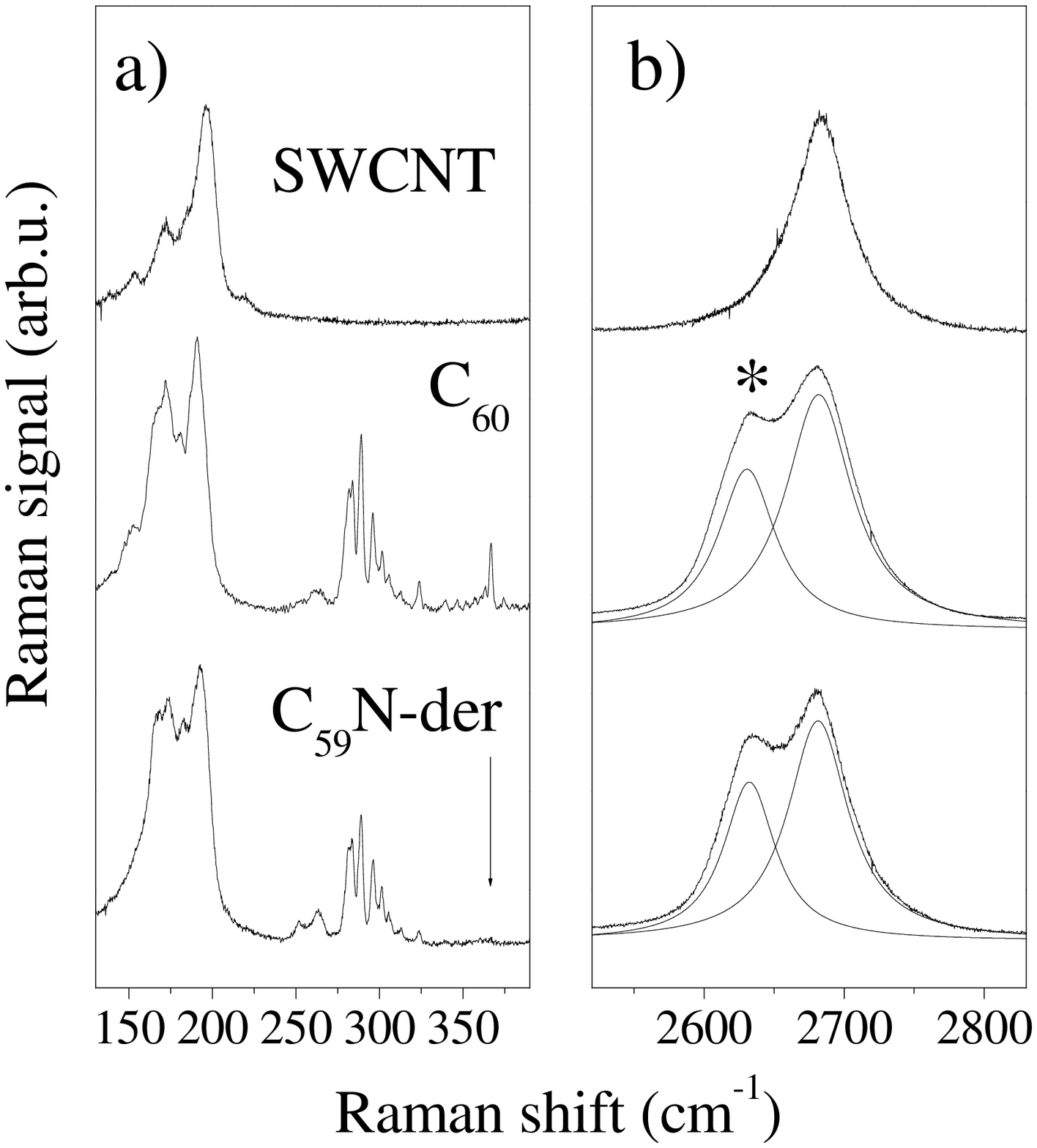}
\caption{Raman spectra of the DWCNTs made from C$_{60}$ and
$_{59}$N-der peapods with a) 647 nm and b) 515 nm laser excitation.
The spectra of the starting SWCNT material is shown for comparison.
Arrow indicates a small diameter RBM\ mode that is missing from the
C$_{59}$N-der based DWCNT sample. Solid curves show the
deconvolution of the G' mode spectra into inner (marked by an
asterisk) and outer tube mode components on b). } \label{DWCNTs}
\end{figure}

Encapsulating an azafullerene inside SWCNTs provides a unique opportunity to
explore the in-the-tube chemistry with a heteroatom. Peapod samples with
toluene encapsulated C$_{59}$N-der and C$_{60}$ were subject to 1250 $%
^{\circ }$C dynamic vacuum treatment for 2 hours. In Fig.
\ref{DWCNTs}, we show the Raman response of the resulting DWCNT made
from C$_{60}$ and the C$_{59}$N-der peapod samples. The radial
breathing mode (RBM)\ range \cite{DresselhausFullerenes} is shown in
Fig. \ref{DWCNTs}a at 647 nm laser excitation and the Raman G' mode
range in Fig. \ref{DWCNTs}b with the 515 nm laser. For both
materials, the narrow Raman lines in the 250-350 cm$^{-1}$ spectral
range correspond to the RBMs of inner tubes
\cite{PfeifferPRL2003,SimonDWCNTReview} and the lower frequency G'
mode component (marked by an asterisk in Fig. \ref{DWCNTs}b.) to the
inner tube G' mode \cite{PfeifferPRB2005}. The emergence of the
inner tube RBMs and G' mode in the C$_{59}$N-der peapod material is
a further proof that the azafullerenes are indeed encapsulated
inside SWCNTs. The G' mode is one of the most energetic vibrational
modes of SWCNTs. Therefore any effect related to the modification of
the phonon energies would markedly manifest for this mode. We
observe no change for the inner tube G' mode for the C$_{59}$N-der
based DWCNT as compared to the C$_{60}$ peapod based DWCNTs. Thus,
our result suggests that either nitrogen is not entering to the
inner tube walls or it has no observable effect on the Raman
spectrum.

The only difference observed for the C$_{59}$N-der based DWCNT is the
absence of some modes, corresponding to very small diameter inner tubes
(shown by arrow in Fig. \ref{DWCNTs}a.). This can be understood by the
somewhat larger effective size of the C$_{59}$N-der compared to C$_{60}$ and
by its inability to enter in very small diameter SWCNTs. Since these tubes
remain unfilled, they do not have an inner tube after the heat-treatment.

\section{Conclusion}

We presented the preparation of fullerene peapods containing a derivative of
the azafullerene C$_{59}$N with a low temperature synthesis method. The
encapsulation efficiency of the azafullerene is the same as that of C$_{60}$
fullerenes. Although no spectroscopic evidence for nitrogen on the inner
tube walls was observed, our material is a starting point to explore the
in-the-tube chemistry with heteroatoms. The material might find applications
in nano-electronics as the presence of the sizeable electric dipole moment
of the molecule allows to fine-tune the nanotube properties. Currently, we
investigate the possibility of removing the sidegroup by thermal treatment
to produce C$_{59}$N monomer radicals embedded inside SWCNTs.

\begin{center}
\textbf{Acknowledgement}
\end{center}

Work supported by the FWF project 17345, by the Deutsche
Forschungsgemeinschaft (DFG), by the Hochschuljubil\"{a}umsstiftung
der Stadt Wien project H-175/2001, and by the MERG-CT-2005-022103 EU
grant. FS acknowledges the Hungarian State Grants (OTKA) No.
TS049881, F61733 and NK60984 and the Zolt\'{a}n Magyary fellowship
for support.

$^{\ast }$ Corresponding author: ferenc.simon@univie.ac.at

%\bibliographystyle{nature}
%\bibliography{HabilNano}

\end{document}